\documentclass[namedreferences]{SolarPhysics}
\usepackage[optionalrh]{spr-sola-addons} 
\usepackage{epsfig}                     
\usepackage{graphicx}                    
\usepackage{color}                       
\usepackage{url}                         
\usepackage{rotating}
\usepackage{multirow}

\begin{document}

\begin{article}

\begin{opening}

\title{Magnetic Field Configuration Models and Reconstruction Methods for Interplanetary Coronal Mass Ejections}
\author{N.~\surname{Al-Haddad} $^{1,2}$, T.~\surname{Nieves-Chinchilla}$^{3,4}$, N.~P.~\surname{Savani} $^{5,6}$, C.~\surname{M{\"o}stl} $^{7,8,9}$, K.~\surname{Marubashi}$^{10,11}$, M.~A.~\surname{Hidalgo}$^{12}$,  I.~I.~\surname{Roussev} $^{2}$, S.~\surname{Poedts} $^{1}$, C.~J.\surname{Farrugia} $^{13}$}

%
\runningauthor{Al-Haddad et al}
\runningtitle{CDAW-II: CMEs Magnetic Field Fitting and Reconstruction Codes}
%
\institute {$^{1}${Centrum voor Plasma-Astrofysica, Katholieke Universiteit Leuven, Celestijnenlaan 200B, 3001 Leuven, Belgium}
$^{2}${Institute for Astronomy, University of Hawaii, 2680 Woodlawn Dr., Honolulu, HI 96822, USA.}\\
$^{3}${Heliospheric Physics Lab. GSFC-NASA, Greenbelt, MD. USA.}
$^{4}${IACS-CUA, Washington, DC, USA.}
$^{5}${UCAR, Boulder, CO, USA}
$^{6}${NASA Goddard Space Flight Center, Greenbelt, MD, USA.}
$^{7}${Space Science Laboratory, University of California, Berkeley, CA, USA.}\\
$^{8}${Space Research Institute, Austrian Academy of Sciences, Graz 8042, Austria.}\\
$^{9}${Kanzelh\"ohe Observatory-IGAM, Institute of Physics, University of Graz, Universit\"atsplatz 5, A-8010, Graz, Austria.}\\
$^{10}${Korea Astronomy and Space Science Institute, Daejeon, 305-348, Republic of Korea.}
$^{11}${660-73 Mizuno, Sayama, Saitama 3350-1317, Japan.}\\
$^{12}${SRG-UAH, Alcala de Henares, Madrid, Spain.}\\
$^{13}${Space Science Center and Department of Physics, University of New Hampshire, Durham, NH, USA.}}

%
%

\begin{abstract}
This study aims to provide a reference to different magnetic field models and reconstruction methods for interplanetary coronal mass ejections (ICMEs). In order to understand the differences 
in the outputs of those models and codes, we analyze 59 events from the Coordinated Data Analysis Workshop (CDAW) list, using four different magnetic field models and reconstruction techniques; force-free fitting \cite{Goldstein:1983,Burlaga:1988,Lepping:1990}, magnetostatic reconstruction using a numerical solution to  the  Grad-Shafranov equation \cite{Hu:2001}, fitting to a self-similarly expanding cylindrical configuration \cite{Marubashi:2007} and elliptical, non-force free fitting \cite{Hidalgo:2003}. The resulting parameters of the reconstructions for the 59 events are compared statistically, as well as in selected case studies. The ability of a method to fit or reconstruct an event is found to vary greatly: the Grad-Shafranov reconstruction is successful for most magnetic clouds (MCs) but for less than 10$\%$ of the non-MC ICMEs; the other three methods provide a successful fit for more than 65$\%$ of all events. The differences between the reconstruction and fitting methods are discussed, and suggestions are proposed as to how to reduce them. We find that the magnitude of the axial field is relatively consistent across models but not the orientation of the axis of the ejecta. We also find that there are a few cases for which different signs of the magnetic helicity are found for the same event when we do not fix the boundaries, illustrating that this simplest of parameters is not necessarily always well constrained by fitting and reconstruction models. Finally, we look at three unique cases in depth to provide a comprehensive idea of the different aspects of how the fitting and reconstruction codes work. 
\end{abstract}

\end{opening}
\section{Introduction} \label{intro}
Coronal mass ejections (CMEs) are large-scale eruptions of plasma and magnetic flux with a typical size of  0.25  AU at 1 AU  (e.g., see \opencite{Bothmer:1998}). In the interplanetary space, they typically move at supersonic speeds  between 200 and 2000~km~s$^{-1}$, propagating from the solar corona to 1 AU in a few days. They can interact with Earth's magnetosphere resulting in geomagnetic storms \cite{Gold:1959}. 
Once CMEs drive through and interact with the interplanetary medium, they are often referred to as interplanetary CMEs (ICMEs). ICMEs, measured {\it in situ}, for example at L1, may be composed of a fast forward shock, a dense sheath and ejecta material. Slow CMEs typically do not drive a shock. Hereafter, we use the term ICME once it is propagating in the interplanetary medium after leaving the corona (more or less after 10--20~$R_\odot$). 

ICMEs, measured {\it in situ}, are typically characterized by low proton temperature, the presence of bi-directionally streaming electrons, unusual charge states of oxygen and iron and various magnetic signatures (e.g. \opencite{Lynch:2003,Zurbuchen:2006, Richardson:2010}). However, there is no single characteristic that is consistently observed. Lists of ICMEs measured by ACE and STEREO are maintained by \inlinecite {Jian:2006}, \inlinecite{Lepping:2006}, \inlinecite{Richardson:2010} and \inlinecite{Gopal:2008}, among others. ICMEs have been classified into three different classes based on their magnetic field and plasma properties \cite{Zurbuchen:2006}: 

1. Magnetic clouds (MCs): characterized by low plasma beta, low proton temperature and featuring a strong magnetic field with a smooth rotation \cite{Burlaga:1981}. In order to explain the last characteristic, \inlinecite{Goldstein:1983} suggested that the magnetic field in a MC can be described by a force-free flux rope structure, where $\nabla \times \bf{B} = \alpha \bf{B}$. \inlinecite{Burlaga:1988} took a constant $\alpha$, in which case the flux rope satisfies the Lundquist solution \cite{Lundquist:1950} and can be expressed in terms of Bessel functions. This became the most widely-used model to describe the magnetic field structure of MCs. MCs are believed to represent about one third of the total number of ICMEs \cite{Gosling:1990,Richardson:2010}. However, a number of other studies proposed that a larger fraction of ICMEs are MCs; for example, the study by \inlinecite{Li:2011} and \inlinecite{Marubashi:2000} estimated the proportion of MCs to be 50\% and 80\%, respectively. Other authors have studied the variation of the proportion of MCs among ICMEs during the solar cycle \cite{Kilpua:2011,Richardson:2010} and found that it varies from close to 100$\%$ near solar minimum to about 20-25$\%$ near solar maximum.

2. Ejecta  or irregular and weak magnetic field ICMEs: These configurations have some but not all the properties of MCs. Typically, their magnetic field is too weak and/or irregular or without much rotation. Some of these ejecta can be further classified into sub-categories, for example, ``magnetic cloud-like'' \cite{Zhang:2007} or magnetic flux rope \cite{Rouillard:2010}. It has been recently proposed that some of these events are actually MCs whose legs passed through the observing spacecraft \cite{Marubashi:2007}, which has been confirmed with STEREO observations \cite{Moestl:2010}. According to this type of studies, ejecta are MCs but are not observed as such due to an observational bias.

3. Complex ICMEs (Complex ejecta), which may result from the interaction of successive CMEs or from the interaction of CMEs with complex solar wind structures and streams \cite{Burlaga:2002,Wang:2003,Lugaz:2005b}. Double rotation within one MC have also been reported from {\it Ulysses}  \cite{Rees:2004} and ACE observations \cite{Steed:2011}, showing that even isolated ICMEs may produce complex ejecta.

Understanding the nature of the magnetic field in ICMEs (magnetic clouds and ejecta) is a crucial part of comprehending their behavior and the way they affect the interplanetary medium and Earth' s magnetosphere throughout their propagation. Thus there has been several attempts to describe the structure of the magnetic field in ICMEs using models and magnetic field reconstruction codes. Figure~1 gives some examples of proposed models.

\begin{figure*}[t*]
\begin{center}
{\includegraphics*[width=10cm]{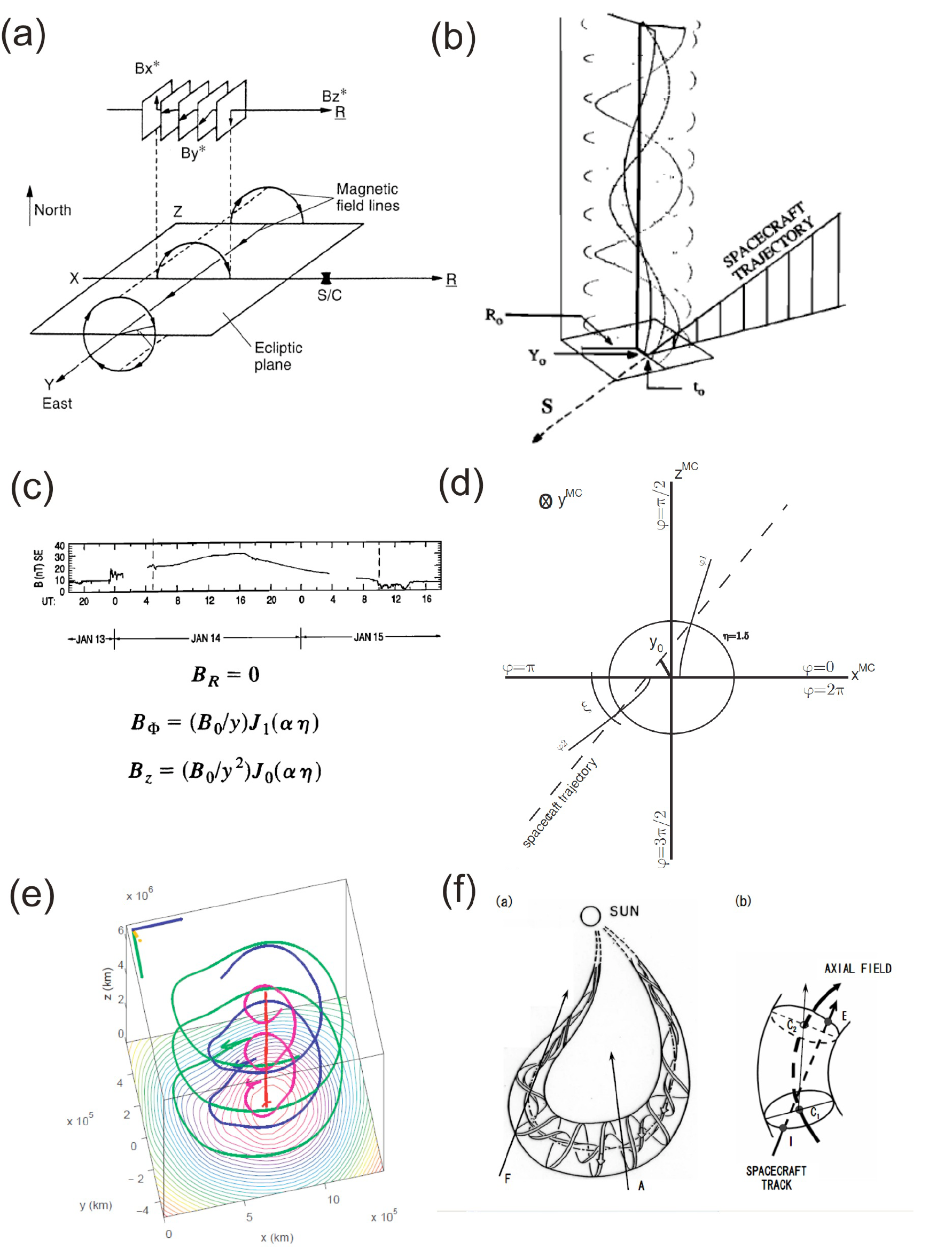}} 
\end{center}
\caption{An  overview of different models to understand MCs.
(a) Goldstein (1983) proposed that MCs are force-free
structures, which was confirmed by Marubashi (1986) (figure adapted from Bothmer and Schwenn,1998)
(b) Lepping {\it et al.} (1990) succeeded in fitting the Lundquist model to {\it in situ} magnetic field
observations after an idea by Burlaga (1988). 
(c) Farrugia {\it et al.} (1993) included expansion into the model.
(d) Schematic representation of the spacecraft trajectory inside of a MC of ellipitical cross-section following the model of Hidalgo {\it et al.}, 2002.
(e) Hu and Sonnerup (2001) model
magnetic flux ropes using magneto-hydrostatic reconstruction technique without a predefined geometry. 
(f) Marubashi and Lepping (2007) include curvature into the classic model.}
\end{figure*}

In order to explain the behavior of the magnetic field in MCs, early models were developed to accommodate \inlinecite{Burlaga:1981}'s definition of a MC. Most of these models restrict the internal structure of the magnetic field to the force-free configuration where the magnetic field, {\bf B}, is described as  $\nabla \times {\bf B}  = \alpha {\bf B}$. Several magnetic field reconstruction and fitting models have been built upon this definition; among the most used is the model by \inlinecite{Burlaga:1988} and \inlinecite{Lepping:1990} which assumes a cylindrically symmetric solution and a constant $\alpha$ across the cloud.

\inlinecite{Marubashi:1986} also adopted a force-free model, but without assuming a constant $\alpha$. \inlinecite{Farrugia:1993}  and \inlinecite{Farrugia:1995} introduced a cylindrical model incorporating self-similarly expansion in two initially force-free constant-$\alpha$ codes, a cylindrical and a spherical models, and noted that the cylindrical configuration did not maintain the force-free state after it starts to expand. Thereafter, \inlinecite{Shimazu:2002} provided a modification of the mathematical formalism in order to keep the self-similarly expanding cylindrical model as force-free. 

Several other models have emerged thereafter which make different assumptions: fitting to a cylinder of  elliptical cross section by \inlinecite{Hidalgo:2002} and \inlinecite{Hidalgo:2003}; fitting to a kinematically distorted flux rope \cite{Owens:2006a,Vandas:2003,Vandas:2006,Demoulin:2009}; non-cylindrical flux rope fitting \cite{Mulligan:2001,Owens:2012}; torus fitting \cite{Romashets:2003,Marubashi:2007}. 
The Grad-Shafranov reconstruction technique \cite{Hu:2002,Moestl:2009} assumes a structure in magneto-hydrostatic (MHS) equilibrium with an invariant direction, and uses the Grad-Shafranov equation to describe the magnetic field in the structure. Therefore, this model does not only use the magnetic field measurements but also measurements of the plasma pressure, and it is only applicable to a structure possessing axial symmetry.  The Grad-Shafranov technique has been tested using multi-spacecraft measurements \cite{Liu:2008b,Kilpua:2009a,Moestl:2009d} and, recently, improvement to the algorithm have been presented by \inlinecite{Isavnin:2011}. 
 
While most techniques reconstruct or fit ICMEs as some type of twisted flux ropes, this result is not necessarily well tested. It has recently been shown by \inlinecite{AlHaddad:2011} that the Grad-Shafranov technique is designed such that it will always reconstruct  a helical flux rope from a rotating magnetic field observed by a single-spacecraft, even if the magnetic field is not a helical flux rope in 3-D. This is precisely because Grad-Shafranov reconstruction assumes an invariance along the cloud axis (so-called 2.5 D method). As was shown in \inlinecite{AlHaddad:2011}, this is equivalent to  assuming a helical solution from the beginning. The assumption of invariance along the cloud axis is also made in all the other fitting techniques used in this article. Other recent observations pointing towards a more complicated geometry of ICMEs at 1~AU have been made by \inlinecite{Kahler:2011} and \inlinecite{Farrugia:2011}. 

Using 2-D and 3-D numerical simulations, it is possible to show that a CME initiated at the solar surface (for example using a twisted flux rope) evolves through its interaction with the solar wind into a typical ICME \cite{Riley:1997,Riley:2003,Odstrcil:1999,Manchester:2004b,Chane:2006,Shen:2007}. This idea has now been confirmed by STEREO observations (see, for example, \opencite{Savani:2010}). In these simulations, synthetic satellite measurements made at 1~AU show a typical ICME structure with a fast shock, preceding a dense sheath and an ejecta. Synthetic coronagraphic and heliospheric images are also able to reproduce typical views of CMEs \cite{Lugaz:2005a,Manchester:2008,Riley:2008,Lugaz:2009b,Odstrcil:2009}.

\inlinecite{Riley:2004} made a comparison of magnetic field reconstruction and fitting models for ICMEs, by fitting two different time series of a simulated ICME to five different techniques: three force-free models, the elliptical model by \inlinecite{Hidalgo:2002} and the Grad-Shafravov reconstruction code. The results of the reconstructions were then compared back to the 2.5-D MHD simulation. One of the difficulties that the authors encountered was choosing the boundaries. Another one was recognizing a MC within the provided time series. Specifically, when the minimum variance technique is used and the impact parameter is large, it is hard to observe the properties characterizing a MC. 

Three of the models used in the study by \inlinecite{Riley:2004} are also used in the present study. We also added the self-similarly expanding cylindrical fitting of \inlinecite{Marubashi:2007} and only kept one classical --no expansion-- force-free model.  However, it is important to stress that in this project we apply the comparison to a list of observed events. Here, the authors address some of the issues discussed in \inlinecite{Riley:2004} by performing more detailed analyses and proposing fixed boundaries.

In this study, we compare the reconstruction of four different magnetic field fitting techniques for most of the fast ICMEs and MCs observed during solar cycle 23, for which a source region can be identified (see details in section \ref{models}). Our goals are to acquire a broader understanding of the magnetic field structure in MCs and to investigate if there are statistical differences between the codes for MCs and non-MC ejecta, and take steps towards deciphering if all ICMEs can be expressed as flux ropes if a sophisticated enough model is used. We do this by comparing the results of the different fitting techniques. 
In section \ref{models}, we describe the data used in this study followed by a succinct description of the different codes. In section \ref{results}, we discuss the full results for the 59 selected events and in section \ref{case}, we present two events for which we used the same boundaries for all the codes. We conclude and discuss our results for the nature of magnetic fields in ICMEs in section \ref{conclusion}.

\section{Data and Models Overview} \label{models}

\subsection{CDAW Data}
The data is taken from the list of Coordinated Data-analysis Workshop (CDAW) of MCs and ejecta during solar cycle 23. The list can be found at
\url{http://cdaw.gsfc.nasa.gov/meetings/2010_fluxrope/LWS_CDAW2010_ICMEtbl.html}. This list contains all shock-driving ICMEs from solar cycle 23 for which it was possible to determine the source region, and for which the source region was within $\pm 15^\circ$ of disk center.  This list is a subset of the list of all interplanetary shocks observed during solar cycle 23 from \inlinecite{Gopal:2010}. The list contains 59 ICMEs classified as follows: 24 events as Magnetic Clouds (MCs) and 35 events as Ejecta (EJ).

\subsection{Grad-Shafranov Reconstruction Technique}

For the Grad-Shafranov (GS) reconstruction, we use the code by \inlinecite{Hu:2002} and followed the guidelines for using it discussed by \inlinecite{Moestl:2009}. This is based on magneto-hydrostatic equilibrium of a system with an invariant direction and is a solution for what is basically a numerical boundary problem. The GS equation is:

\begin{equation}
    \frac{\partial^2 A}{\partial x^2} +\frac{\partial^2 A}{\partial y^2} = -\mu_0 \frac{d P_t(A)}{d A},
\end{equation}

where $A(x,y)$ is the magnetic vector potential, $P_t$ the transverse pressure defined as $P_t=p+B_z^2/(2 \mu_0)$ with $p$ being the thermal pressure. The total magnetic field is given by $\bf{B} = \nabla A \times {\bf e_z} + B_z {\bf e_z.}$
The first step to solve this equation is to determine the invariant axis, $z$. This is done first using a minimum variance analysis on the magnetic field components and by finding a frame where the transverse pressure  is a single-value function of the magnetic vector potential, $A$. After this, the GS equation can be solved numerically. 

The major assumptions of the technique are: the structure is assumed to be invariant along the ICME axis (so called 2.5 D), and it is assumed to be time-independent during the whole measurement by the spacecraft. Compared to other techniques, it does not include expansion or a toroidal geometry. One of the potential issues with the Grad-Shafranov technique is the stability of the integration and the influence of the solver and stabilization procedure on the results. However, Grad-Shafranov reconstruction has been tested by using observations of ICMEs by two spacecraft at a significant separation with respect to the ICME size \cite{Liu:2008b,Moestl:2009,Moestl:2009d}. Some of the stability issues have also been recently addressed in \inlinecite{Isavnin:2011}.

One of the main advantages of the GS method is that the shape of the flux rope's cross-section and the number of flux ropes inside the reconstructed interval are both not pre-defined, but are an output of the technique. Another advantage is that the boundaries of the ICMEs are chosen through an optimization procedure to make the $P_t(A)$ function as close to single-valued as possible. The $P_t(A)$ function is fitted with a polynomial function with exponential tail(s) and only these events for which the fitting residue is small (typically less than 0.1) and the fitting appears visually correct are deemed as successfully reconstructed. Further details can be found in \inlinecite{Hu:2004}.   

GS reconstruction gives the following results: flux rope orientation (longitude and latitude), impact parameter (closest approach of the spacecraft to the MC axis), magnetic flux (axial and poloidal), axial current, the cross--section shape and the central field strength. 
The GS reconstruction feasibility was checked for every event in the CDAW list.  We used primarily data from the Wind spacecraft (MFI \cite{Lepping:1995} and SWE \cite{Ogilvie:1995} instruments), and switched to ACE (MAG \cite{Smith:1998} and SWEPAM \cite{McComas:1998} instruments) if there was a data gap at Wind. The GS reconstruction method was successful for:

\begin{itemize}
\item full list: 20 events out of 59 (34 \%)
\item 17 MC events out of 24  (71 \%)
\item 3 EJ events out of 35 (8 \%)
\end{itemize}

This means that the applicability of using GS reconstruction is closely tied to the definition of a magnetic cloud (higher-than-average field strength, smoothly rotating magnetic field vector, low $\beta$, low $T_p$). Typically, this means that, for these cases, one magnetic field component ($B_y$ or $B_z$) needs to be bipolar and the other unipolar, depending on the orientation of the flux rope (see discussion in \opencite{Bothmer:1998}). Also, we were able to reconstruct about 1/3 of the events in the complete list, which is similar to the often quoted ratio between all ICMEs and those which contain parts that  satisfy the MC definition.

\begin{figure*}[t*]
\begin{center}
{\includegraphics*[width=12 cm]{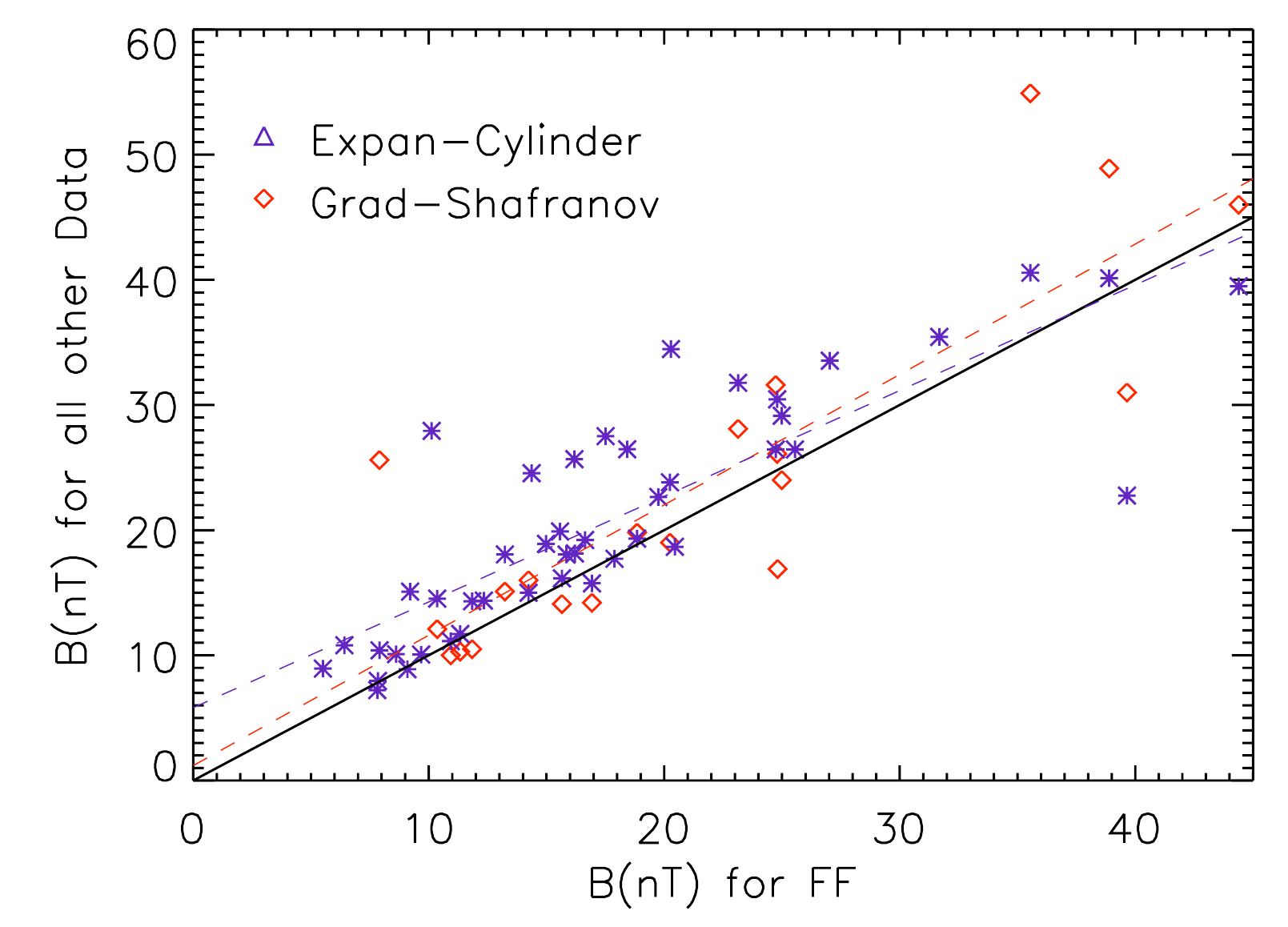}} 
\end{center}
\caption{Axial magnetic field for the Grad-Shafranov (red diamonds) and self-similarly expanding cylinder (purple stars) models as compared with the axial magnetic field for the FFCA model. The red and purple dashed lines show the value of the best-linear fit to the Grad-Shafranov and self-similarly expanding cylinder models, respectively.} 
\end{figure*}



\subsection{Self-similarly Expanding Cylinder Fitting Technique}

This model is a force-free model that assumes a spatially constant $\alpha$ and performs the fitting in cylindrical coordinates. Here, $\alpha$ is chosen fixed for every event and equal to 2.4. The model takes into account self-similarly expansion, as was initially suggested by \inlinecite{Farrugia:1992,Farrugia:1993}. The original inclusion of the self-similarly expansion has been found to cause a deviation from the force-free state during expansion. Some refinements proposed by \inlinecite{Shimazu:2002} have been incorporated into the model to preserve the force-free state for the duration of the MC travel. As a result, the magnetic field intensity $B$ and the radius $r$ of the fitted MC at a given time as produced by the model are defined as follows: 
\begin{equation}
    {\bf B} = \frac {B_0} {(1+ \frac{t}{t_\mathrm{exp}})^2}
\end{equation}

\begin{equation}
    {\bf r} =  {r_0} {(1+ \frac{t}{t_\mathrm{exp}})}
\end{equation}

where $B_0$ is the magnetic field intensity on the axis of the CME cylinder, $t_\mathrm{exp}$ is the expansion time, $t$ is the time since the first encounter of the spacecraft with the CME, $r_0$ is the distance between the cylinder axis and the spacecraft at the time of the first encounter.

\begin{figure*}[t*]
\begin{center}
{\includegraphics*[width=10cm]{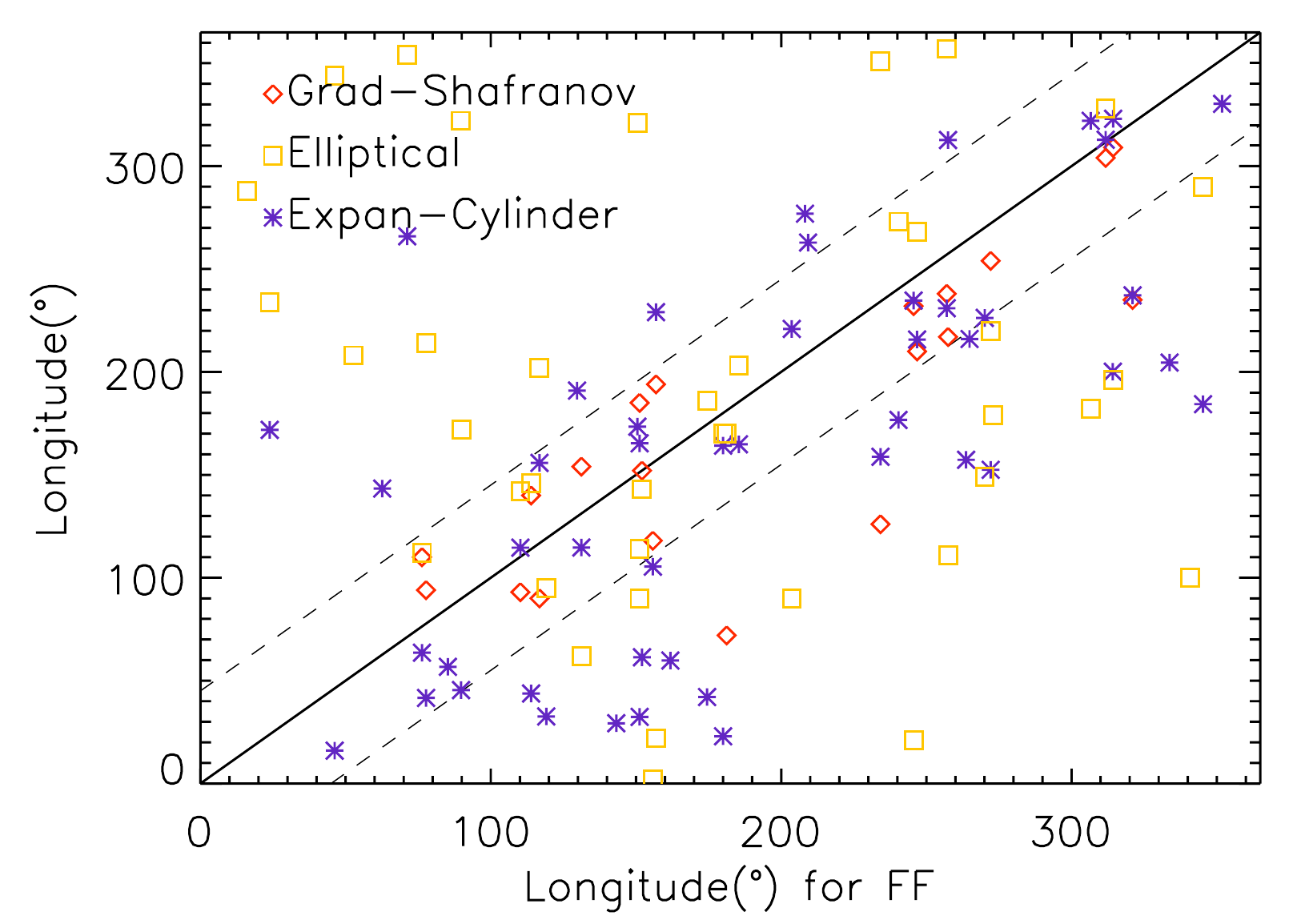}} 
{\includegraphics*[width=10cm]{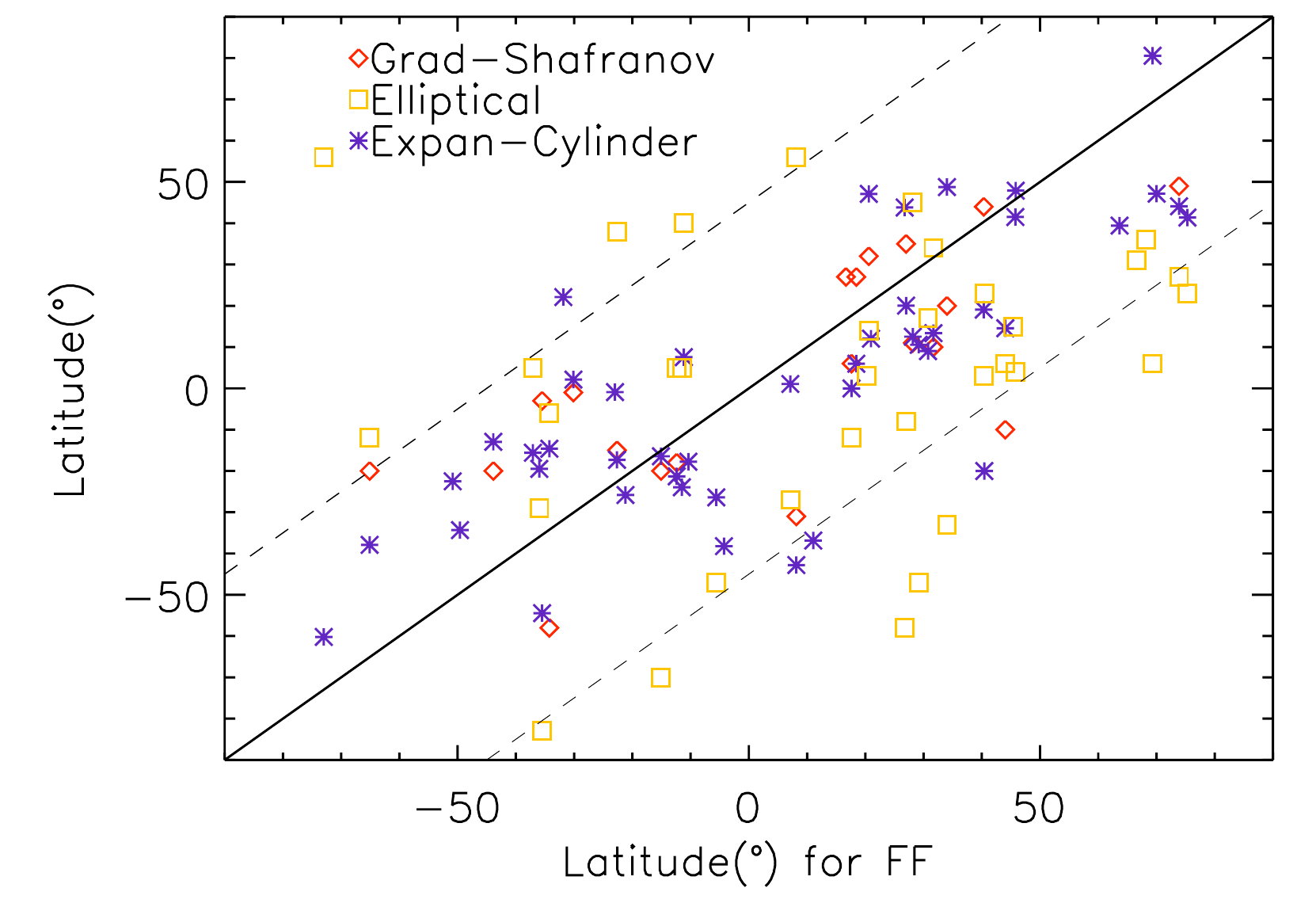}}
\end{center}
\caption{Same as Figure~2 but for the longitude and latitude of the axis of the reconstructed ICME. The yellow squares are for the elliptical cross-section fitting model. The dashed lines highlight to angles $\pm 45^\circ$ from the FFCA angle.} 
\end{figure*}

The model also provides values for the longitude and latitude of the axis, as well as the impact parameter, which is defined to as the distance between the closest axis of cylinder and the path of the spacecraft. The model fits simultaneously the following quantities for both chiralities (positive and negative): $B_0$, $r_0$, $t_\mathrm{exp}$, the longitude, latitude and impact parameter. The final selection is made by choosing the sign of the chirality, which gives the best fit. The details of the model can be found in Appendix A of \inlinecite{Marubashi:2007}. The self-similarly expanding cylinder fitting method was successful for:

\begin{itemize}
\item full list: 50 events out of 59 (85\%)
\item 22 MC events out of 24  (91.5 \%)
\item 28 EJ events out of 35 (80 \%)
\end{itemize}

There were seven events, which were found to be contracting (negative $t_\mathrm{exp}$) and three events, for which the expansion time was greater than 500 hours, which correspond to less than 10$\%$ change in the axial magnetic field strength over the crossing of a typical magnetic cloud ($\sim$ 1 day). 

\subsection{Force-Free Constant-Alpha  Reconstruction}

This force-free fitting model is based on the  force-free flux constant-alpha (FFCA) rope developed by \inlinecite{Burlaga:1988} and then optimized by \inlinecite{Lepping:1990}.  In our model, $\alpha$ is fixed to equal 2.4. This implies the flux rope's magnetic field on the outer shell is almost completely poloidal. The value of 2.4 is not optimized and it could be changed (i.e. reduced in value) if it is believed that on the outer shell of the flux rope the field lines are still helical. This might be the case if the rear or front edge of the flux has been ``eroded'' away from its ideal case (perhaps by reconnection) as proposed by \inlinecite{Dasso:2004} and recently confirmed by \inlinecite{Ruffenach:2012}. The $\alpha$
parameter has also been used as a free variable is some models (e.g. \opencite{Lepping:1990}). Since $\alpha$ is constant and fixed to equal 2.4 for the FFCA and for the self-similarly expanding cylinder models, it is clear that these two models are identical when the expansion is equals to 0.

The flux rope axis is calculated by Minimum Variance Analysis (MVA). MVA is a technique originally developed for solving the normals to a tangential discontinuity (TD) \cite{Sonnerup:1998}.
 In order to obtain the orientation of an ICME, the axis is 
determined by the intermediate eigenvector, as pointed out by \inlinecite{Goldstein:1983}. 
 
 
The chirality of the flux rope is determined by the sense of direction the magnetic field rotates in the MVA reference frame. The impact parameter and magnetic field 
magnitude are  the only two free variables. They are optimized using a non-linear optimization routine \cite{Nelder:1965}. For each event, the fitting was attempted with two different boundaries: those from \inlinecite{Lepping:2006} and those selected by K. Marubashi. In addition, for eight events, a different set of boundaries were chosen directly by N.~P. Savani as the other two sets of boundaries were judged as not satisfactory for these events. Overall, the method was successful for:
 
\begin{itemize}
\item full list: 55 events out of 59 (93\%)
\item 24 MC events out of 24  (100\%)
\item 31 EJ events out of 35 (88.5\%)
\end{itemize}

\subsection{Elliptical Cross-section Fitting Model}

The elliptical, non-force-free model \cite{Hidalgo:2002,Hidalgo:2003} considers that ICMEs have elliptical cross-section due to their interaction with the solar wind. In this model, the expansion of the local cross-section is taken into account.  Two characteristics distinguish this model from the others used in this paper: 1) non-force-free condition is assumed, and, 2) an elliptical cylindrical coordinate system is chosen to solve the Maxwell equations.
The model assumes a flux rope magnetic field geometry and treats the radial and axial electric currents separately. The local solutions of the Maxwell equations provide the three magnetic field components to be transformed into spacecraft coordinates. The only assumptions are that the radial and axial components of the current density are constant. The cross-section of the flux ropes is not required to be circular. This model was compared with the circular cross-section model of \inlinecite{Lepping:1990} in \inlinecite{Nieves:2005} and compared with information from remote-sensing measurements in \inlinecite{Nieves:2012}.

The ellipticity, the current density components, expansion time of the cloud and the axial magnetic field are determined by fitting the magnetic field data as explained in section 2 of \inlinecite{Hidalgo:2003}.  Multiple regression analysis is used to infer the spacecraft trajectory through the flux rope. The method converged for all events. However, the fit goodness was reported with a flag: 0=bad; 1=good; and, 2=very good.  Such ``goodness'' is based on the correlation coefficient, fitting residue and visual goodness. In the rest of the analysis, we take into account just fitting results with quality 1 or 2 (i.e., removing bad fits). In that case, the elliptical cross-section fitting method was successful for:

\begin{itemize}
\item full list: 39 events out of 59 (66 \%)
\item  24 MC events out of 24  (100 \%)
\item 15 EJ+ events out of 35 ( 43\%)
\end{itemize}

\subsection{Selection of the ICME's Boundaries}

Selection of the boundaries of an ICME is still a challenging problem. This is because there is no single definitive characteristic of ICMEs. Some authors base the ICME boundaries purely on magnetic properties; others use low proton temperature as one of the criteria; while others use charge state information. Many of the {\it in situ} signatures have been taken into account in order to determine the boundaries of ICMEs. However, each of those indicators provide different boundaries which leaves the problem still unsolved.
In this study, we let each group determine the boundaries for the ICMEs. For three events, we re-run the analysis using pre-set boundaries as given by the self-similarly expanding cylinder fitting technique. 


\section{Comparative analysis of the fitting parameters -- Results} \label{results}


Out of 59 events on the CDAW list, the reconstruction of the magnetic field was successfully done for 20 events, mostly MCs, by the GS technique. By comparison, the FFCA fitting and the self-similarly expanding cylinder methods were able to fit more than 85\% of the list, and the  elliptical cylinder fitting method 66\%. In the following sections, we compare the fitted parameters from different codes with that from the FFCA code because (i) it is the simplest code in our study, and (ii) it succeeded in fitting most of the cases with the least deviation from the results of the other models. The number of cases that have been analyzed in common between the FFCA code and the other codes is as follows: 20 with the GS reconstruction technique, 45 with the self-similarly expanding cylinder fitting method, and 37 with the elliptical cylinder fitting method.

\subsection{Chirality}

For all 20 events reconstructed by the GS code, the chirality (sign of the magnetic helicity) was the same as for the self-similarly expanding cylinder and the FFCA model. There are three events for which the FFCA and self-similarly expanding cylinder found opposite signs for the helicity  (13, 15 and 52). 
This shows that even the simplest result of fitting and reconstruction methods is not necessarily well constrained. The difference is primarily due to the choice of boundaries. In two of these three cases (13 and 15), when the FFCA fit is performed again with a different choice for the boundaries, it returns the same sign of helicity as that found by the self-similarly expanding cylinder. 
In fact, it is worth noting that the chirality found with the FFCA differs for 7 events depending on whether the boundaries are chosen following \inlinecite{Lepping:2006} or that chosen specifically for this study by K. Marubashi. 

\begin{figure*}[ht*]
\begin{center}
{\includegraphics*[width=5. cm, height = 6.8cm]{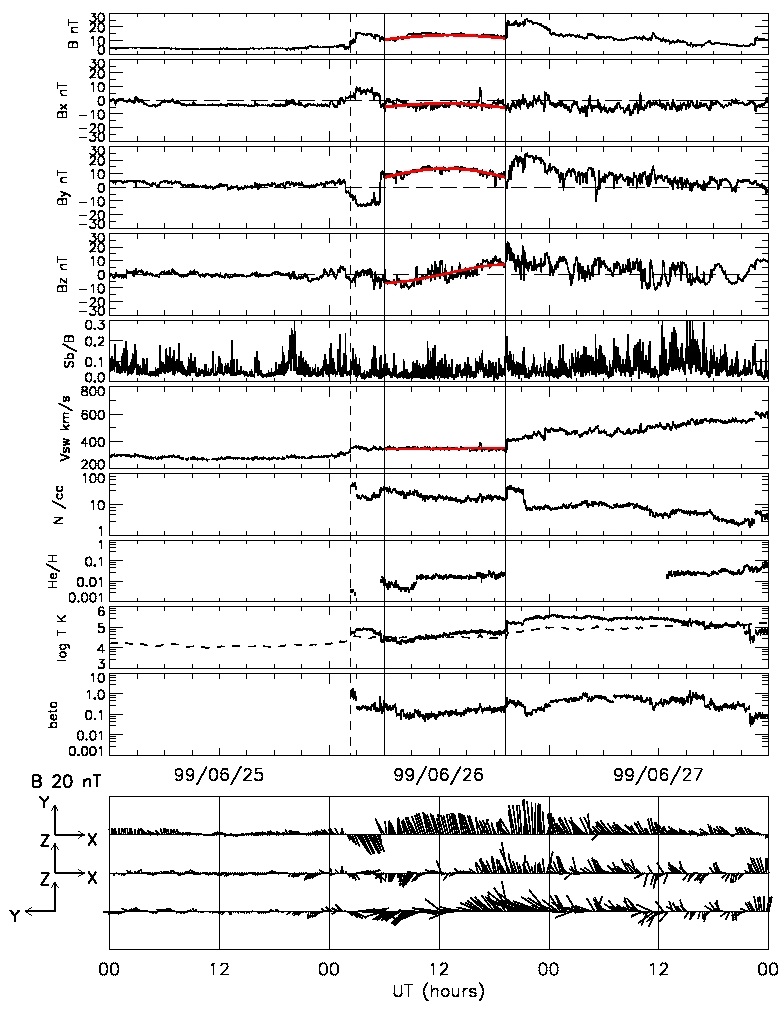}} 
{\includegraphics*[width=5.5 cm]{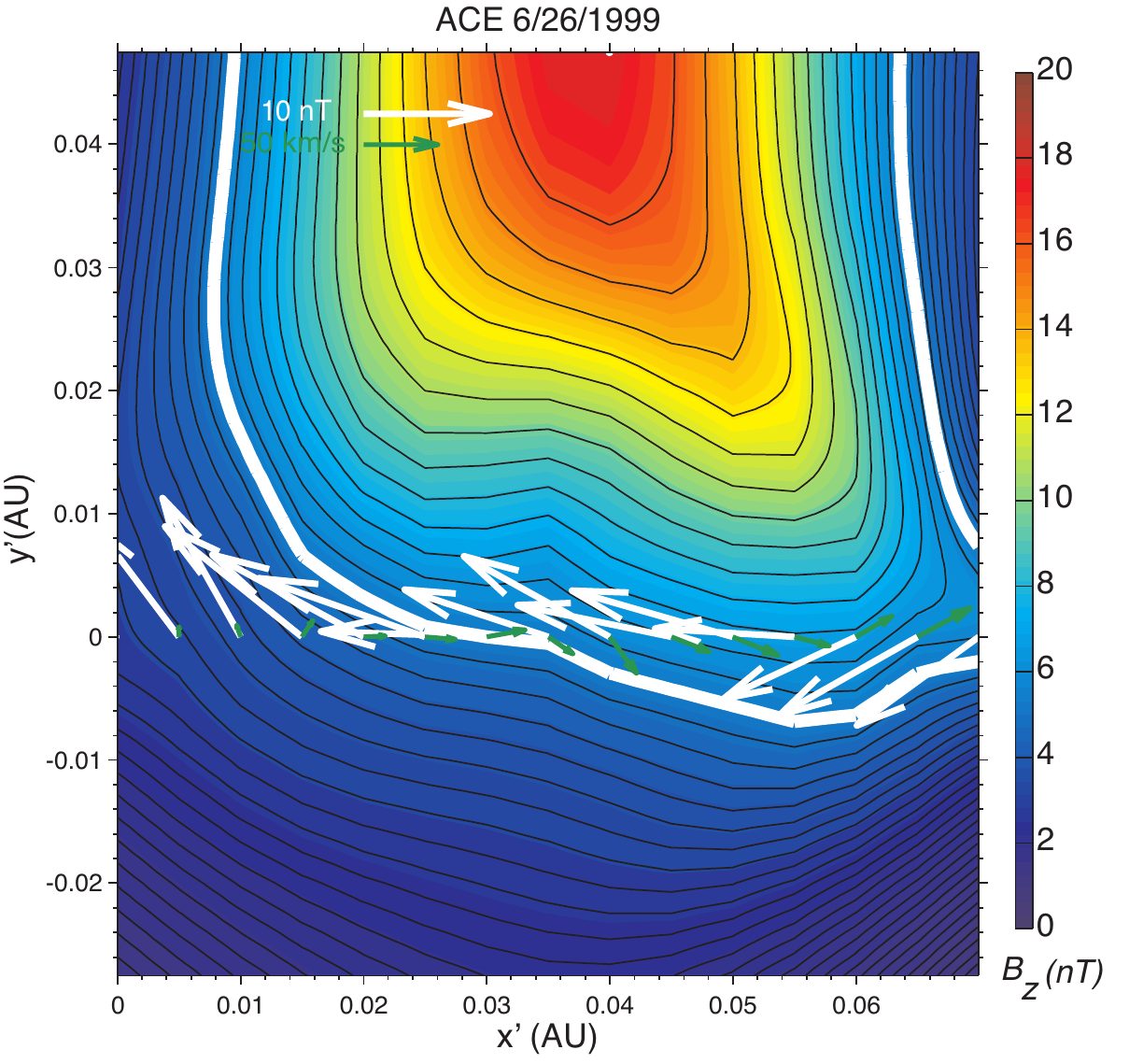}} 
{\includegraphics*[width=5.5 cm, height = 7.5cm]{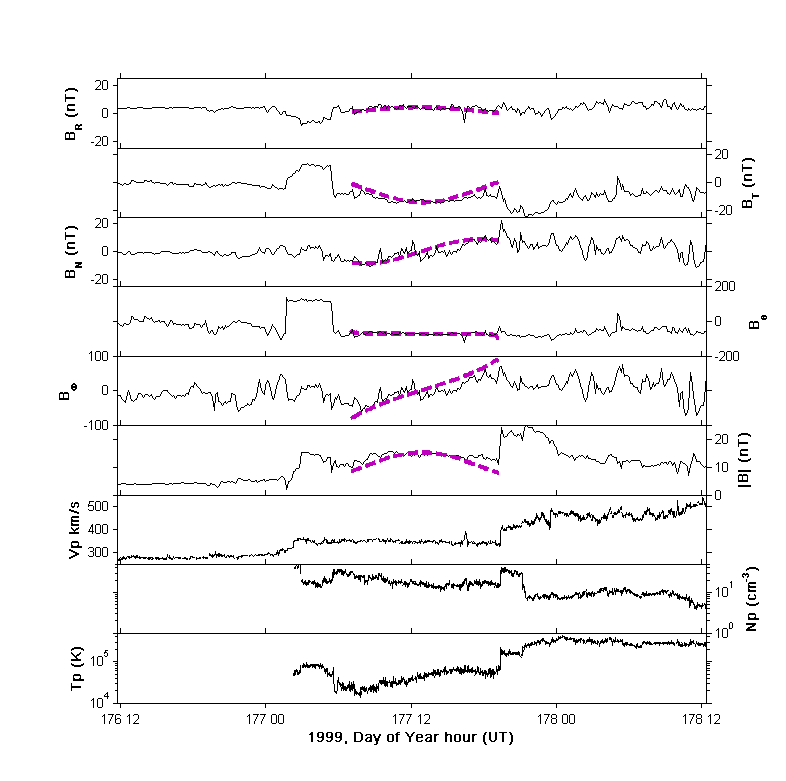}}
{\includegraphics*[width=5 cm,  height = 7.5cm]{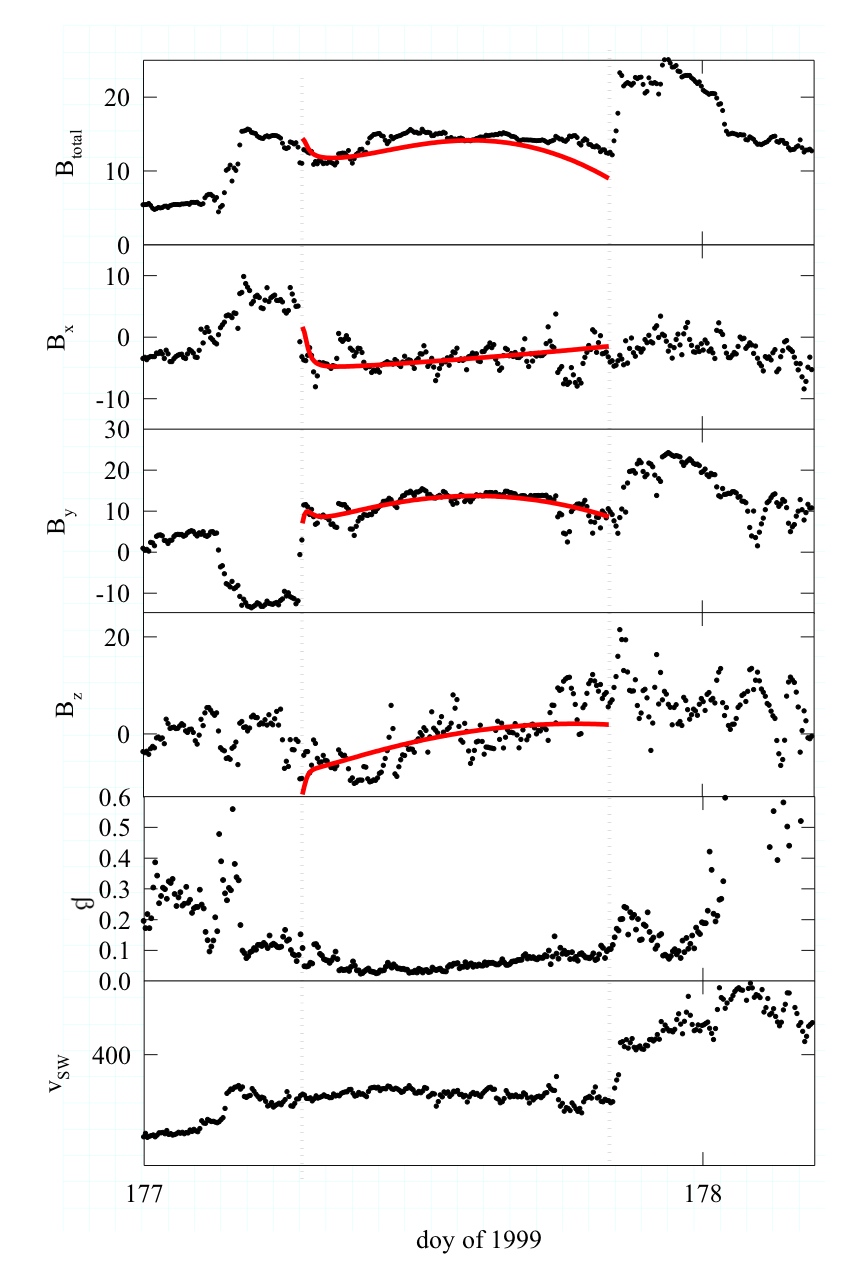}}

\end{center}
\caption{Fitting for event 10a by the different codes. Top left: Self-similar expanding cylinder code fit shown in red. The different panels are the total magnetic field, $B_x$, $B_y$, $B_z$, the magnetic field fluctuation, the solar wind speed, proton density, alpha-to-proton ration, temperature and plasma $\beta$ from top to bottom. Bottom left: FFCA code fit shown in purple. The panels show from top to bottom, the radial, tangential, normal  (R, T and N) components, the longitude and latitude angles (in RTN) and magnitude of the magnetic field, the proton velocity, density and temperature. Top right: GS reconstruction map, the axial magnetic field is color-coded and the black contours are magnetic field lines in the plane of the cross-section of the ICME. Observed magnetic field components in this plane are shown with white vectors and velocity in green. Bottom left: Elliptical cross-section fit in red. The different panels show the magnetic field, its three components, the plasma $\beta$ and the solar wind speed from top to bottom.}
\end{figure*}

One event is more ambiguous. For event 52, the FFCA fit is relatively poor and gives a left-handed ejecta for the boundaries selected by K. Marubashi as well as for those from \inlinecite{Lepping:2006}, whereas the self-similarly expanding cylinder fit gives a right-handed ejecta. It should be noted that the best-fit axial magnetic field by FFCA for this event is less than 6 nT for the two sets of boundaries, reflecting a very weak ejecta.

This overall result about the chirality may appear inconsistent with what was found in previous studies by \inlinecite{Dasso:2003} and \inlinecite{Dasso:2004} where the magnetic flux and helicity was found to be nearly independent of the fitting and reconstruction model, for two well observed MCs. Here, we find that it is generally true, consistent with these results. However, we also show that for ejecta, for which the boundaries are not well defined because of incomplete rotation, fittings with different boundaries can yield opposite chirality for the same event.



\subsection{Maximum Magnetic Field Strength}

For all the common cases between the FFCA code and one of the other codes in this study, we calculated the correlation of the fitted axial magnetic fields. For the self-similarly expanding cylinder fitting method, we use the axial magnetic field halfway through the cloud, corresponding approximately to the magnetic field at the closest approach to the cylinder's axis, as calculated from Equation~(2). We do not compare the value of the axial magnetic field for the elliptical cross-section fitting model with that of the other models. This is because the axial magnetic field in this model is the sum of two fields: the magnetic field created by the currents and the axial field which is a parameter from the model.  Figure~2 summarizes the magnetic field results. 

Overall, the correlation between the FFCA, GS and self-similarly expanding cylinder code is quite good ($\sim$0.84). The axial magnetic field value from the GS code tends to be slightly higher than that from the FFCA especially for stronger magnetic fields, whereas the self-similarly expanding cylinder code tends to return higher values for the magnetic field for weak magnetic fields.

\subsection{Orientation of the ICME Axis: Longitude and Latitude}

The orientation of the ICME as it arrives at Earth is an essential parameter, since it is often used to determine the CME rotation as it propagates in the heliosphere \cite{Yurchyshyn:2001}. In addition, knowing precisely the orientation of an ICME is particularly important since the orientation of an ICME is known to be related to its geo-effectiveness (see, for example \opencite{Zhao:1998}). As can be seen in Figure~3, the orientation of some ICME events differs greatly depending on which model is used. In fact, only for one CME (event $\#$44) did all events give an orientation of the axis within $\pm 45^\circ$ of each other.

First, we focus on a comparison of the orientation between the Grad-Shafranov, self-similarly expanding cylinder and FFCA models for the 20 events successfully reproduced by all three methods. 
For seven events, the three methods give an orientation within $\pm 45^\circ$ of each other. For eleven other events, the Grad-Shafranov reconstruction technique gives an orientation of the axis within $\pm 45^\circ$ of one of the other two methods (but not the other). In fact, although the FFCA and self-similarly expanding cylinder fitting methods are closely related, the axis of the fitted cylinder is found to be in agreement for only about 45\% of the events (21 out of the 45 common cases; for 14 of these, the orientation is within $\pm 30^\circ$). While this may appear surprising considering the similarities between the two methods, it should be noted that the force-free method without expansion relies on the minimum-variance analysis to obtain the orientation of the flux rope, whereas the expanding cylinder method relies on a multi-parameter fitting procedure to minimize the deviation between the model and the measurements. In addition, for many events, both methods used different boundaries.

Comparing the direction obtained from the elliptical cross-section model with that from the force-free cylindrical models (with and without expansion), we find that the direction from all three models only agree for one event. For seven additional events, the direction from the elliptical cross-section model is within $\pm 45^\circ$ of the direction from the FFCA fitting model; for three additional cases, the same is true as compared to the direction from the self-similarly expanding cylinder model, and for one event it is true as compared with the direction for the Grad-Shafranov reconstruction technique. 

Overall, there are only 34 events reconstructed by three methods or more. For 30 of these events, the orientation of the axis of the reconstructed ICME is consistent between at least 2 methods. The same is only true for three out of the 18 events reconstructed by two methods only. Finally, seven events were reconstructed by one method only. 

\begin{figure*}[t*]
\begin{center}
{\includegraphics*[width=5.5 cm, height = 7.5cm]{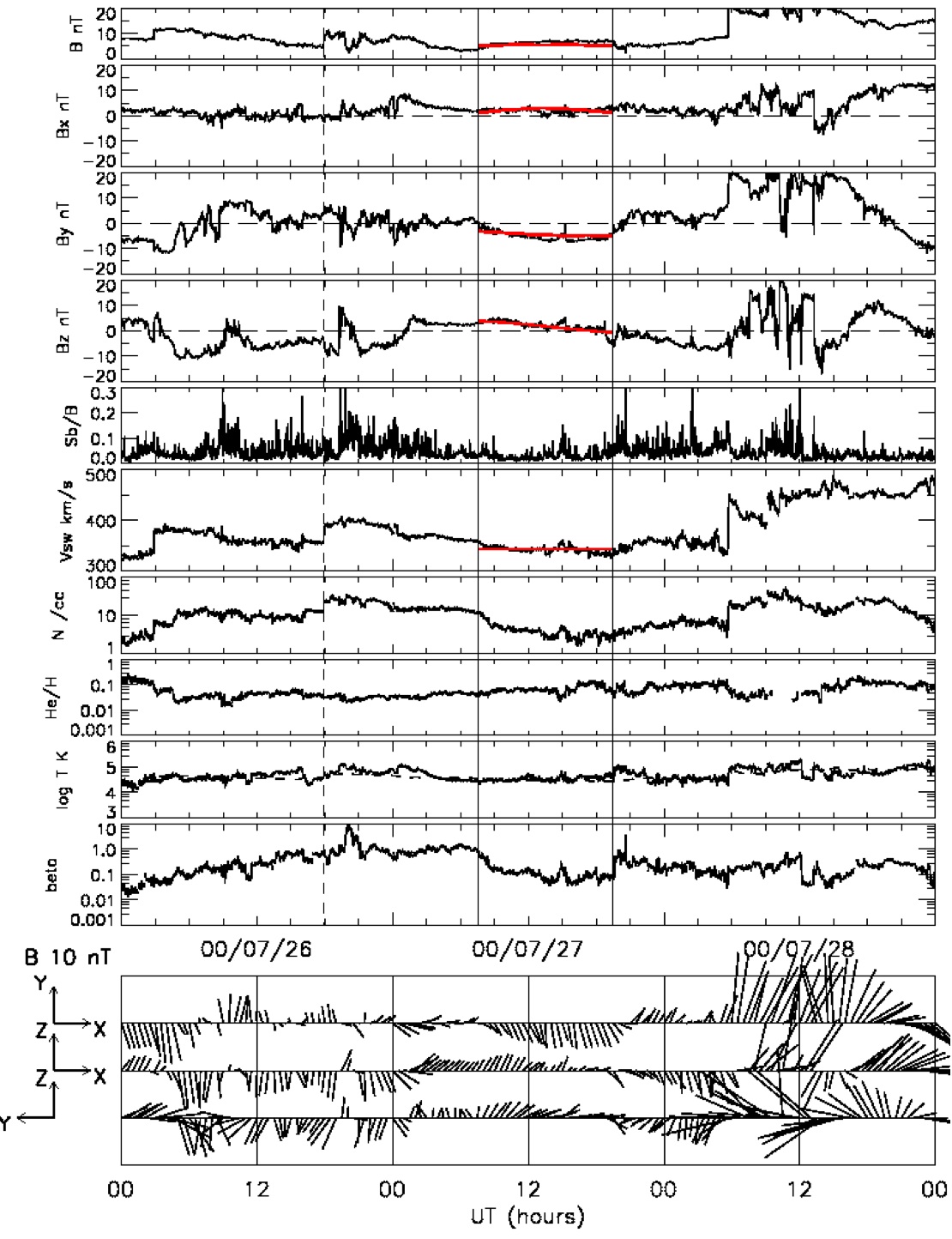}} 
{\includegraphics*[width=5.5 cm]{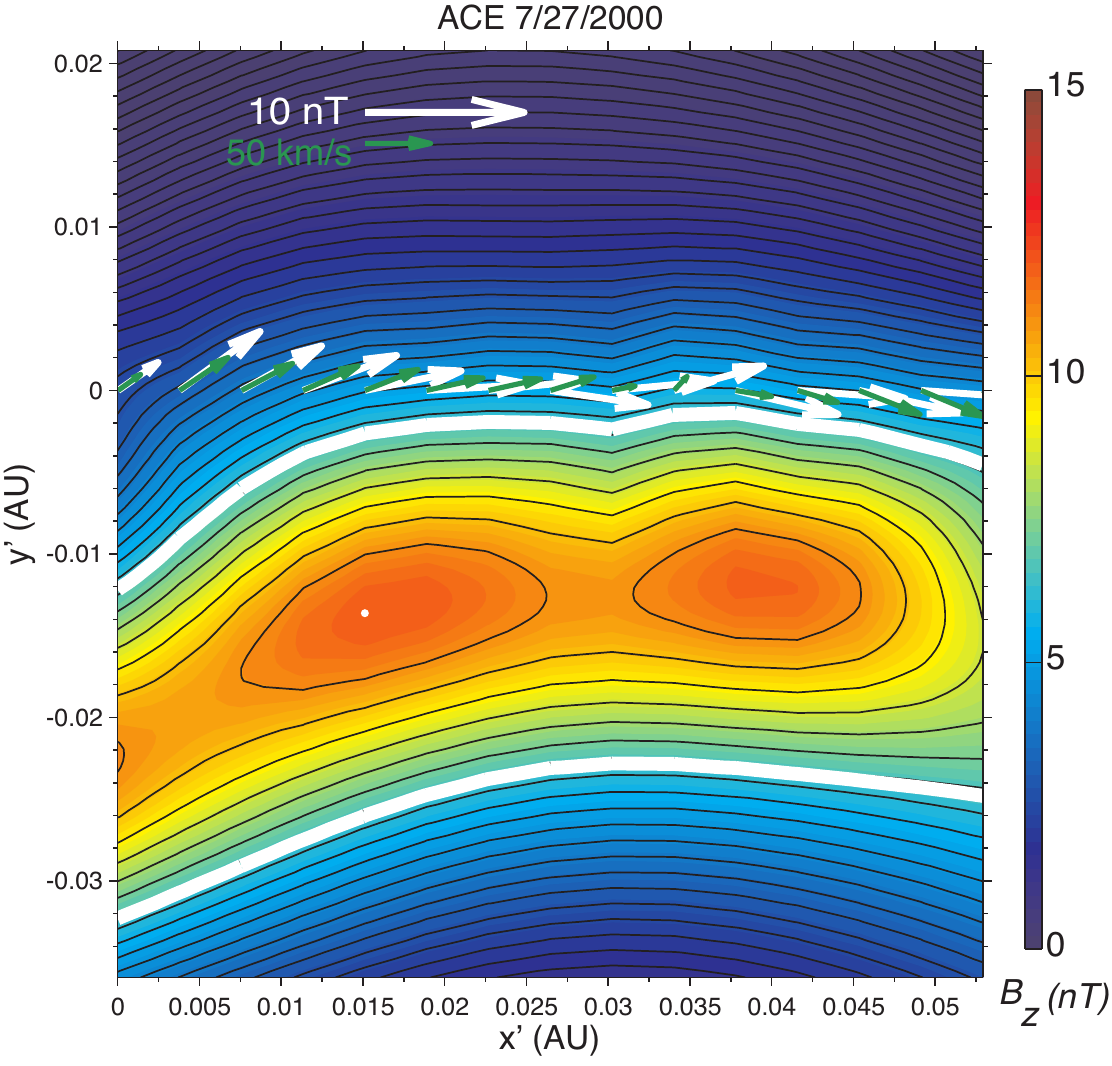}} 
{\includegraphics*[width=5.5 cm, height = 7.5cm]{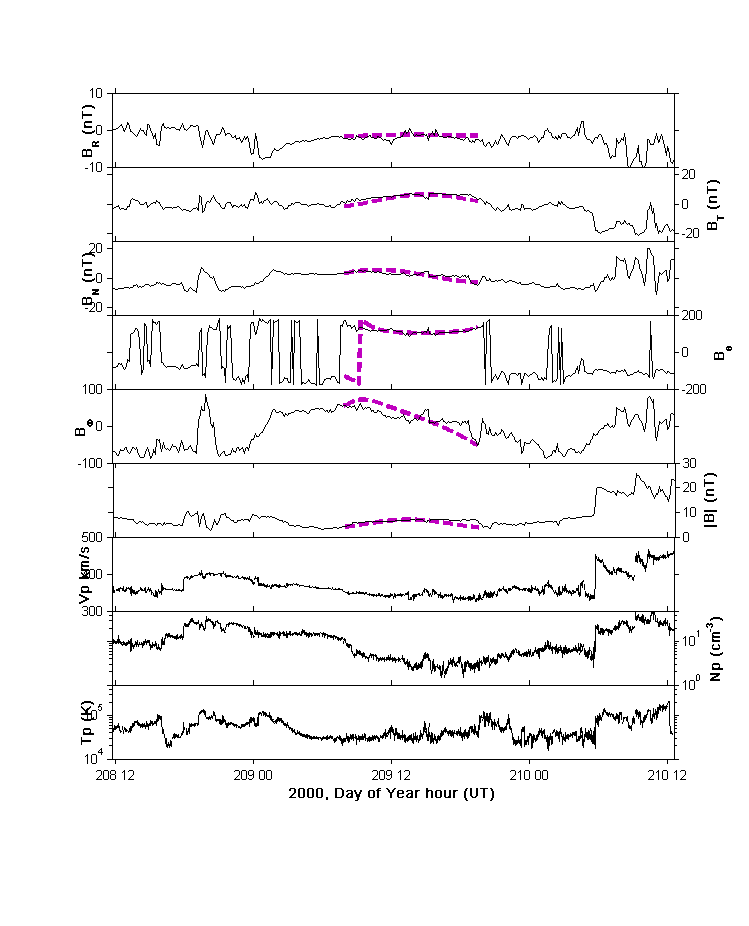}}
{\includegraphics*[width=5 cm, height = 7.5cm]{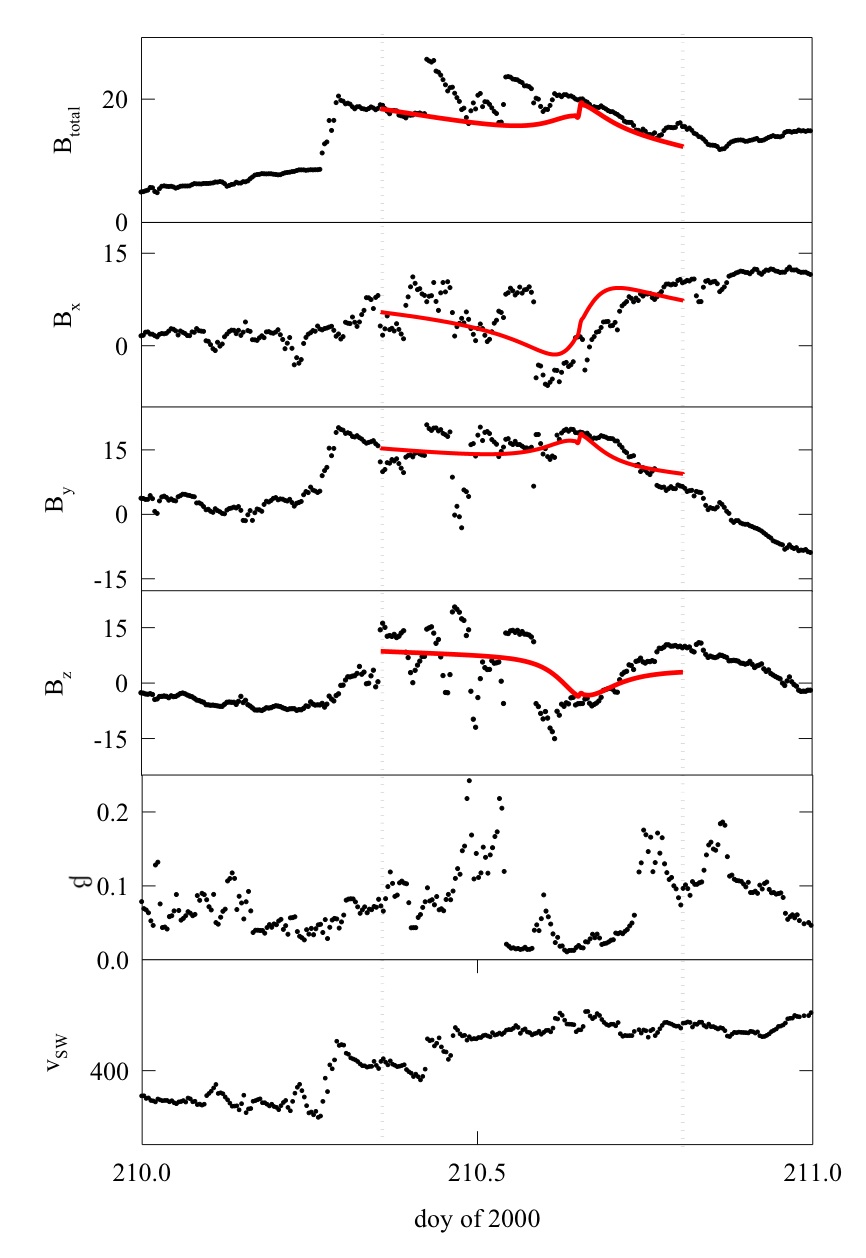}}

\end{center}
\caption{Same as Figure~4 but for event 20.} 
\end{figure*}

\subsection{Difference Between MCs and non-MC Ejecta}

Here, we briefly compare the average values of some parameters for the MCs with those for the non-MC ejecta. We only focus on the FFCA and self-similarly expanding cylinder, because the Grad-Shafranov reconstruction methods is only able to reconstruct three non-MC ejecta (out of 35) and the elliptical cross-section fitting has different set of parameters than these two models.
Consistent with their definitions, non-MC ejecta are found to have, on average, weaker magnetic field than MCs: 13.75 nT vs. 22.1 nT (FFCA) and 17.7 nT vs. 34.4 nT (expanding cylinder). We also find that the average impact parameter is larger for non-MC ejecta than for MCs: 0.73 vs 0.54 (expanding cylinder) and 0.32 vs. 0.29 (FFCA). This may indicate that ejecta are indeed flux rope-like magnetic clouds but they are measured at a large impact parameter, resulting in weaker magnetic field and less well-defined properties. Note that the average impact parameter for the events reconstructed by the GS method is 0.29 (FFCA) and 0.47 (expanding cylinder), once again showing that the GS method was only successful for MCs or MC-like ejecta.

\section{Detailed Analyses of Three Events} \label{case}

We focus on events 10, 20 and 27 which we fitted using the same boundaries. These were those chosen  for the self-similarly expanding cylinder fitting method. 

\subsection{Event 10}

This event is a left-handed non-MC ejecta observed in 1999 June 26-28. This was one of the events resulting in a ``poor'' fit for the elliptical cross-section model and it was not initially successfully reconstructed by the Grad-Shafranov method. For this method, one problem was the presence of a shock propagating inside the ICME at 19:16 UT on June 26. Because the GS reconstruction method attempts to find the best orientation of the ICME by requiring the function $P_t(A)$ to be single-valued, the presence of a shock in the magnetic and plasma pressure in one side of the cloud makes this an impossible task. This has been previously noted in \inlinecite{Kilpua:2009a} and \inlinecite{Moestl:2009d}. It is not a problem encountered by the other fitting methods. 

Initially, some researchers found two possible flux rope intervals: one shortly after the shock wave on June 26 lasting about 16 hours (referred hereafter as 10a) and one corresponding to the initial interval fitted with the self-similarly expanding cylindrical code of 6.5 hours at the end of June 27 and in the beginning of June 28 (referred hereafter as 10b from 20:30 UT on June 27 to 03:00 UT on June 28). We only discuss here the results from the longer, first interval. The intervals found by different researchers are shown in Table~1. 


The analysis was performed again for the interval 10a using for interval June 26 from 06:00 to 19:00. In Figure~4, we show the fitting results for the different methods corresponding to this interval.
Using the same interval for all the methods, the magnitude of the magnetic field in the ICME is found to match relatively well between the two force-free and the Grad-Shafranov  methods and to correspond to a value $\sim 15-20$~nT. The orientation of the cloud is found to be consistent with a latitude of $0^\circ \pm 12^\circ$ and a longitude of 65$^\circ \pm 25^\circ$ between these three methods, whereas the elliptical cross-section model finds a value about 90$^\circ$ away. This example illustrates the importance of the choice of boundaries for the study of magnetic clouds and ICMEs, as three of the four codes find very similar parameters when the same boundaries are chosen.

\begin{table}[ht]
\begin{tabular}{|clc|c|c|}
\hline
{\bf Event} & {\bf 10} (June) & {\bf 20} (July) &{\bf 27} (Nov.)\\

\hline
{\bf GS}  & N/A & 08:36(27)--19:24(27)  & 23:56(06)--17:20(07) \\
{\bf FFCA} & 06:00(26)--22:00(26)& 07:30--19:30 & 22:10(06)--17:10(07) \\
{\bf Exp. Cylind.} & {\bf 06:00(26)--19:00(26)}  & {\bf 07:30--19:30}  &  {\bf 22:10--17:10} \\
{\bf Ellip} & 22:36(26)--02:36(27) &  10:10--20:40 & 23:08--17:45\\
\hline
\end{tabular}
\caption{Intervals used by different researchers for the events 10, 20 and 27. In bold are the intervals chosen when the events were compared with the same boundaries. In parentheses are the day of the month corresponding to the interval boundary. Event 10, 20 and 27 took place in June 1999, July 2000 and November 2000, respectively.}
\end{table}



\begin{figure*}[t*]
\begin{center}
{\includegraphics*[width=5.5 cm, height = 7.5cm]{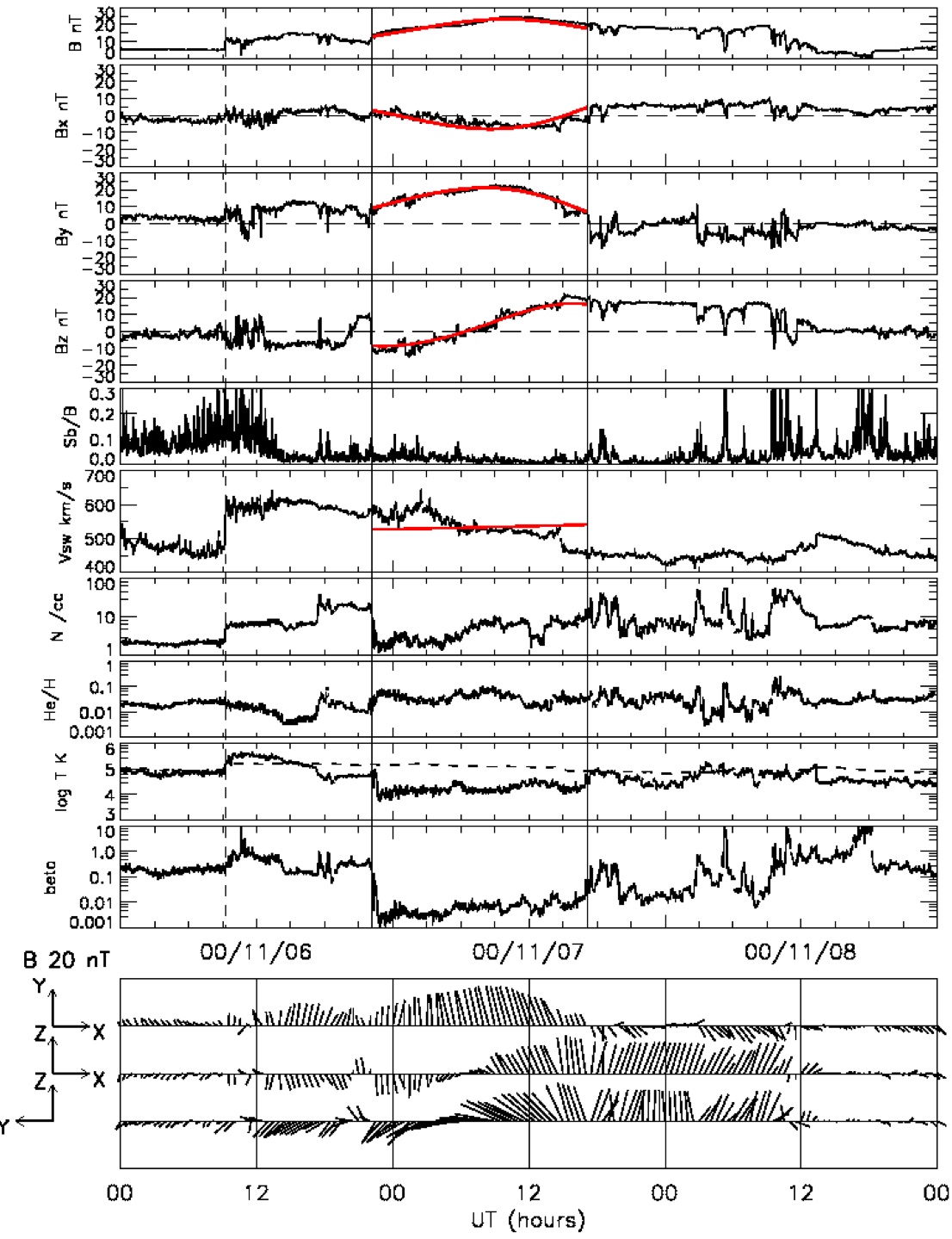}} 
{\includegraphics*[width=5.5 cm]{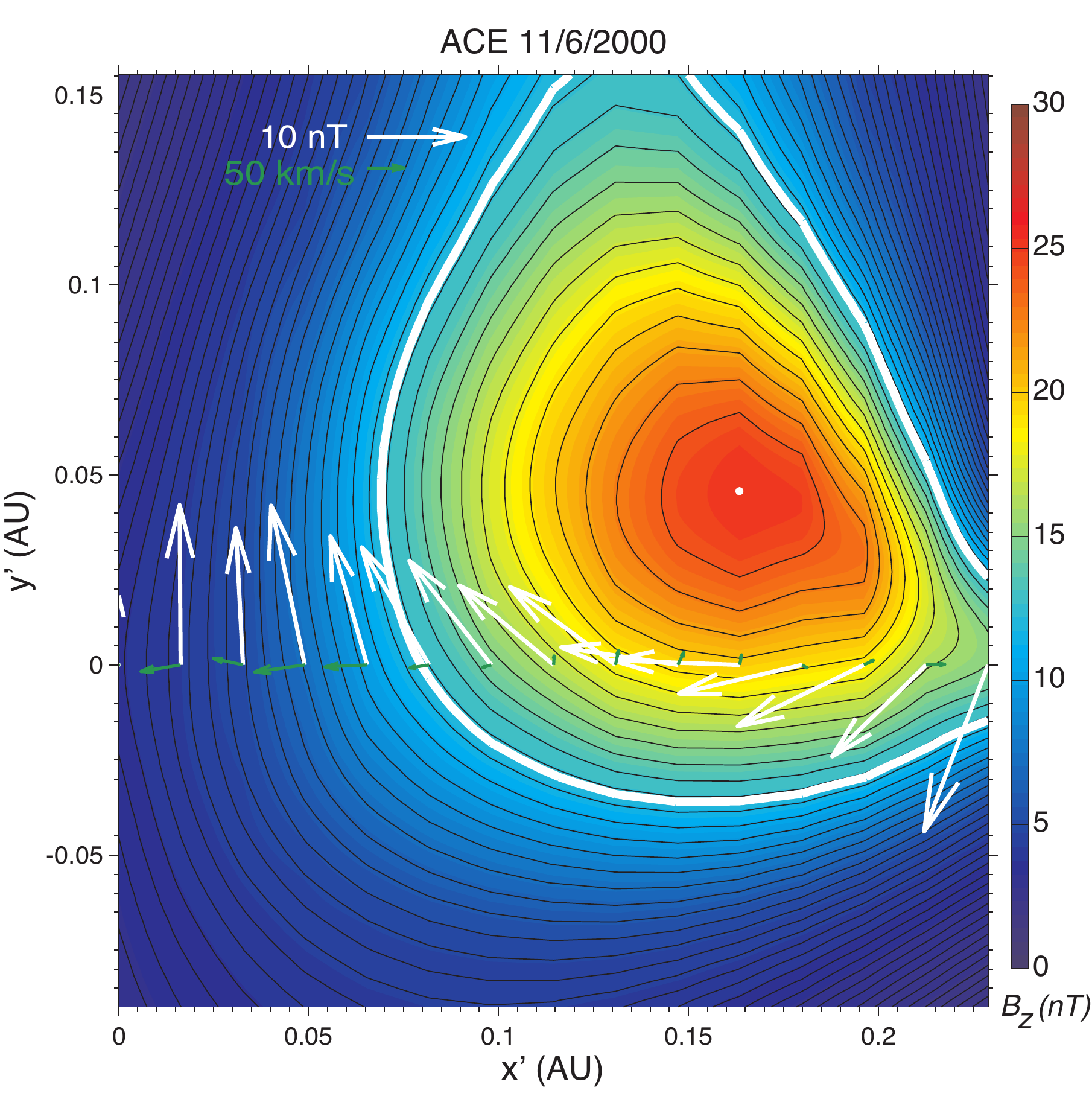}} 
{\includegraphics*[width=5.5 cm, height = 7.5cm]{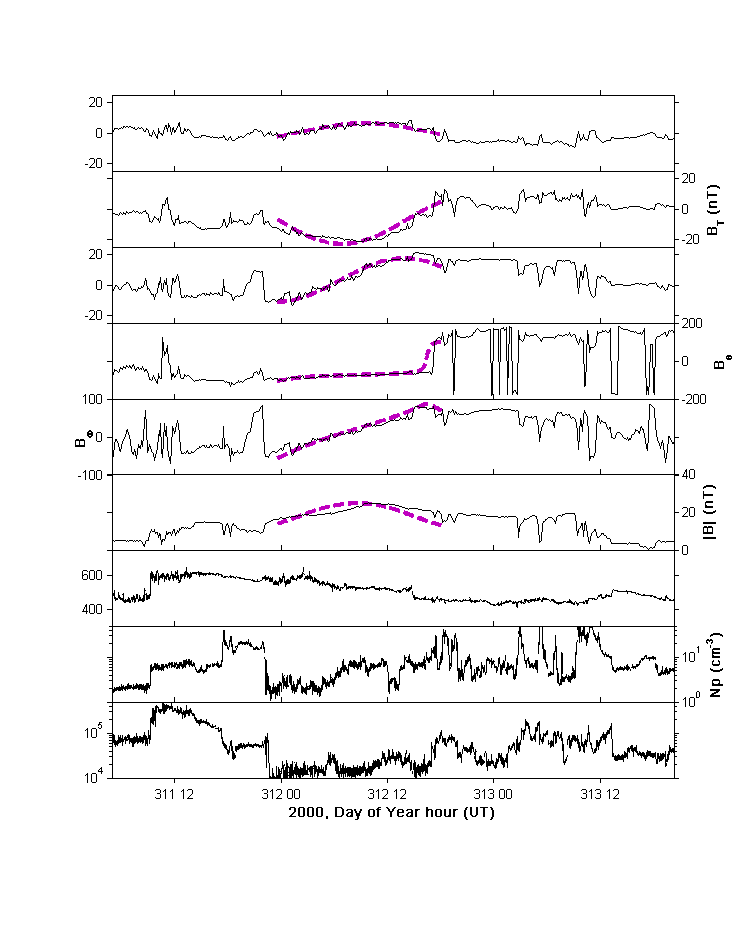}}
{\includegraphics*[width=5.5 cm, height = 7.5cm]{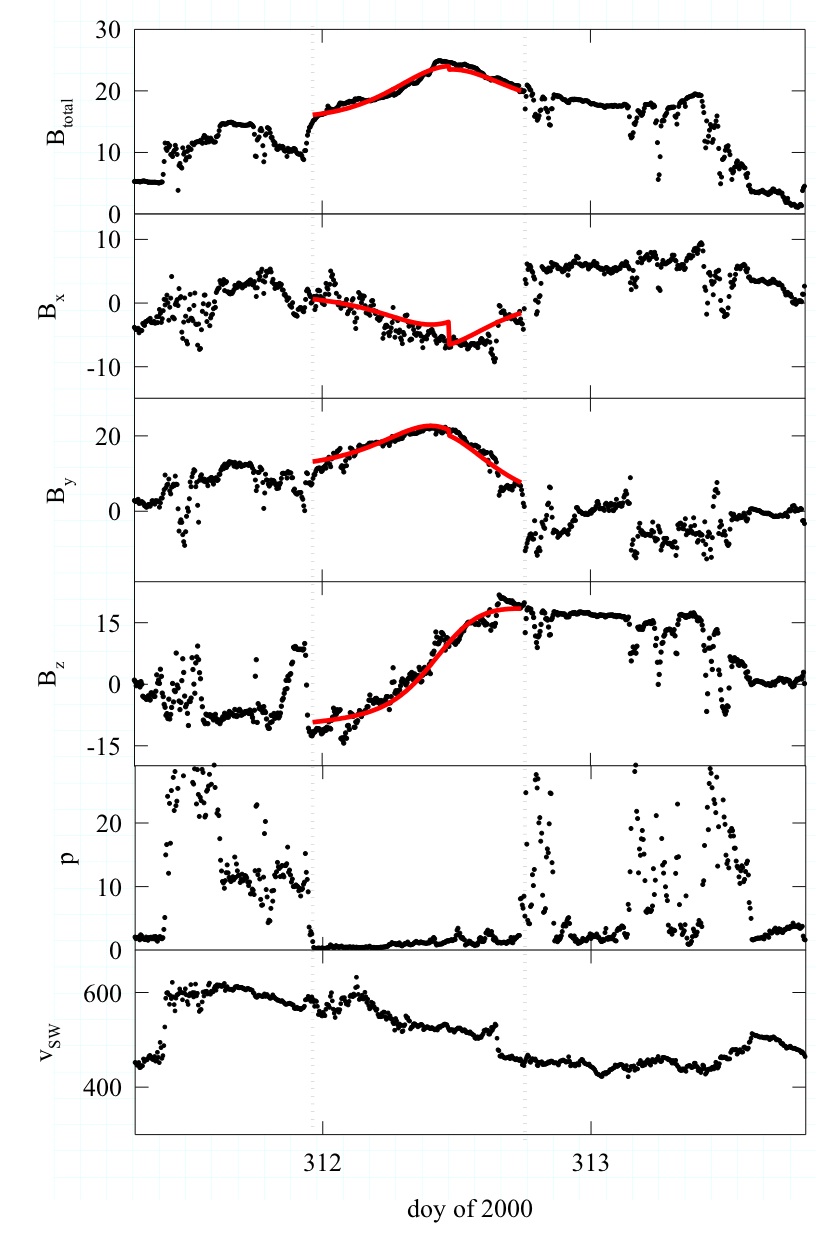}}

\end{center}
\caption{Same as Figure~4 but for event 27} 
\end{figure*}

\subsection{Event 20}

This event is a left-handed non-cloud ejecta observed in 2000 July 27. It is one of the three non magnetic cloud ejecta successfully reconstructed using the Grad-Shafranov reconstruction method. Initially, as we let each researcher select the ejecta's boundaries, the direction of the cloud axis was in good agreement between the four methods: a longitude of 240$^\circ \pm 30^\circ$ and a latitude of 22$^\circ \pm 12^\circ$. All methods were within 45$^\circ$ of each other, except between the self-similarly expanding cylinder and the elliptical cross-section cylinder which were about 51$^\circ$ apart.

The impact parameter was found to be close to 0 for the Grad-Shafranov but larger than 0.5 for the other three methods. The magnitude of the axial magnetic field was about 20 nT for the Grad-Shafranov method and about 8 nT for the two force-free constant-alpha models.

As can be seen from Table~1, all the different researchers chose boundaries very close to each other and choosing exactly the same boundaries does not drastically change the results, except that the maximum axial magnetic field for the Grad-Shafranov model was 12~nT with an impact parameter of about 0.25, more consistent with the values found by the other models. In Figure~5, we show the results of the fit using the boundaries chosen for the self-similarly expanding cylinder fit.

\subsection {Event 27}

This event is a left-handed  magnetic cloud observed in 2000 November  6-7. All methods agree on the maximum magnetic field strength ($\sim 25$ nT for Grad-Shafranov and the FFCA model and $\sim$31.5 nT for the self-similarly expanding cylinder). However, there is initially little agreement regarding the orientation of the magnetic cloud with a latitude of 14$^\circ \pm 14^\circ$ and a longitude of 145$^\circ \pm 55^\circ$ for the four methods. As for the previous event, the boundaries chosen independently by different researchers were relatively consistent. One significant difference is the inclusion or not of a region of strong magnetic field without much rotation at the ``back'' of the MC (until 02:40 on November 8). Previous studies \cite{Dasso:2004,Dasso:2007,Moestl:2008} have discussed how reconnection between a MC and the ambient solar wind during the ICME propagation may result in this type of regions which, actually, belong to the MC. Each researcher decided exactly where to end the MC. Since there is no counterpart flux on the MC front to the flux on the back of the MC, it is probably better not to take this region into account.

Using the same boundaries, there is a better agreement that the magnetic cloud has a low inclination (latitude $\sim 5$) but not much better agreement about the precise orientation or the value of the impact parameter. The fits and reconstructions are shown in Figure~6.


\section{Summary and Conclusions} \label{conclusion}

In this article, we have presented a comparison of the reconstruction results of 59 ICMEs measured {\it in situ} during solar cycle 23 using four different reconstruction or fitting methods. The events our study focused on were chosen because their source region was known and was within $\pm 15^\circ$ of the central meridian. Our data included 24 events identified as magnetic clouds (MCs) following the definition of \inlinecite{Burlaga:1981} and 35 non-MC ejecta. We find that the two force-free techniques (with or without expansion) are able to fit the vice majority of the events ($\geq$85 \%) including more than 75$\%$ of the ejecta. The Grad-Shafranov reconstruction method which assumes magneto-hydrostatic equilibrium and makes use of the plasma pressure is found to only work for clear magnetic clouds, as it was only successful in reconstructing 8$\%$ of the non-cloud ejecta (3/35). In addition, the Grad-Shafranov reconstruction is not able to reconstruct ICMEs into which a shock is propagating as it violates its assumption. However, one of the advantages of this method is the fact that the boundaries of the ICME are a result of the reconstruction, and they do not need to be chosen before performing the fits, as for the other methods. The expanding elliptical cross-section was able to fit with a reasonable result about two-thirds of the events (all magnetic clouds and about 40$\%$ of the ejecta).

We have found that even the ICME chirality (sign of the magnetic helicity), one of  the simplest possible reconstruction parameters, is not necessarily well constrained and opposite chirality can be found for the same event because of differences in the choice of the boundaries. In most cases, the choice of boundaries alone explain the difference between different methods. 

Regarding the maximum magnetic field strength, we have found that, typically, the results between Grad-Sharanov method, FFCA and self-similarly expanding fittings are well correlated and within about 25\%.
The Grad-Shafranov method returns larger values than the FFCA fitting method for strong magnetic fields, whereas the force-free model which includes self-similar expansion returns higher values than the FFCA fitting method for weak magnetic fields. \inlinecite{Gulisano:2005} previously compared the maximum magnetic field at the cloud axis for 20 magnetic clouds using four fitting methods and found that the variability between methods for the same event was significantly less than the variability between events for the same method (see their figure 4).

For only one event, all four methods find an orientation of the ICME axis within $\pm 45^\circ$. Considering only the two force-free fitting methods (with and without expansion) and the Grad-Shafranov reconstruction method, this is the case for about 30$\%$ of the ICMEs (7 out the 20 common cases).
Directly comparing the two force-free constant-alpha codes (with and without expansion), they return ICME axis within $\pm 30^\circ$ in only $\sim 30 \%$ of the cases and within $\pm 45^\circ$ in only $\sim 45\%$ of the cases. While these two methods assume very similar geometries for the ICME, the fitting procedures are very different: minimum variance analysis is used to determine the axis' direction for the FFCA code, whereas a fitting procedure is used for the expanding cylinder method. This is likely to account for the difference in the direction of the ICMEs' axis.
Finally, the elliptical cross-section model typically gives different orientation as compared to the other methods, being in agreement with at least one other method for only about 30$\%$ of the events. These results were found by  letting researchers select their own boundaries.

It should be noted, that when only two methods successfully reconstruct an event, the orientation found by these two methods usually disagree (by more than $\pm 45^\circ$ in 15 out the 18 such cases). When the two force-free constant-$\alpha$ fitting methods and the Grad-Shafranov reconstruction method successfully reconstruct an event, the orientation from the Grad-Shafranov method is within $\pm 30^\circ$ of that from one of the force-free constant-$\alpha$ method in 70\% of the cases (14 out of 20 such events). Combining all four methods, more than 65$\%$ (23/34) of the events reconstructed by three or more methods have at least two methods giving orientation within $\pm 30^\circ$.
It is therefore clear that having multiple methods able to successfully fit or reconstruct the same event gives more reliable results regarding the orientation of the ICME axis. 

We further quantified the importance of the selection of the boundaries by performing the analyses again for three events using the same boundaries for all methods. By using the same boundaries, we find a better agreement between the different codes regarding the ICME magnetic field strength, and, for one event, a better agreement for the ICME orientation.

Finally, for two methods, we have compared the impact parameter and magnetic field strength of MCs and non-MC ejecta. We find that the non-MC ejecta have, as expected, weaker magnetic field but we also find some evidence that the impact parameter is, on average, larger for non-MC ejecta than for MCs. This is a first statistical hint that most (or all) ICMEs observed at 1~AU may in fact, be magnetic flux ropes, but depending on how they impact the observing spacecraft, they may or may not be recognized as magnetic clouds. Having performed fitting and reconstructions for 59 ICMEs with four different codes, we will further investigate the nature of the magnetic fields in ICMEs as well as how to further improve reconstruction methods in the future.

%
 \begin{acks}
We would like to thank the reviewer for his/her very useful comments which helped improve and clarify our study and manuscript.\\
N.~A. and I.~R. were supported by a NSF grant AGS-0639335 (CAREER). This work was initiated at two Coordinated Data Analysis Workshops (CDAWs) held at Predictive Science Inc. in San Diego, CA and at the University of Alcala in Alcala de Henares, Spain with support from NASA and from Spain.
N.~P.~S. was supported by the NASA LWS Jack Eddy Postdoctoral Fellowship Program, administrated by the UCAR Visiting Scientist Program and hosted by the Naval Research Laboratory. 
C.~M. was  supported by a Marie Curie International Outgoing Fellowship within the 7th European Community Framework Programme. The presented work has received funding from the European Union Seventh Framework Programme (FP7/2007-2013) under grant agreement n$^{\circ}$263252 [COMESEP].
C.~J.~F. acknowledges support from NASA grant NNX10AQ29G and NAS5-0313.

 \end{acks}

%
%
\bibliographystyle{spr-mp-sola-cnd}

\end{article}
\end{document}